\documentclass[11pt,a4paper]{article}
\pdfoutput=1
\usepackage{jheppub}
\usepackage{tikz}
\usetikzlibrary{intersections, calc, arrows.meta, bending}
\usepackage{subcaption}
\usepackage{url}
\usepackage{simpler-wick}
\usepackage{bbm}

\DeclareMathOperator{\Tr}{Tr}

\newcommand{\ri}{\mathrm{i}}
\renewcommand{\th}{\theta}

\newcommand{\cob}{\delta}

\newcommand{\hf}{\frac{1}{2}}

\newcommand{\lap}{\Delta}
\newcommand{\bra}{\langle}
\newcommand{\ket}{\rangle}
\newcommand{\la}{\lambda}

\newcommand{\h}[1]{\widehat{#1}}
\newcommand{\bt}{\beta}

\newcommand{\al}{\alpha}

\newcommand{\rt}[1]{\sqrt{#1}}
\newcommand{\cO}{\mathcal{O}}

\newcommand{\cH}{\mathcal{H}}

\newcommand{\cE}{\mathcal{E}}

\newcommand{\id}{\mathbbm{1}}

\makeatletter
\gdef\@fpheader{}
\makeatother
\begin{document}
\title{Doubled Hilbert space in double-scaled SYK}

\author{Kazumi Okuyama}

\affiliation{Department of Physics, 
Shinshu University, 3-1-1 Asahi, Matsumoto 390-8621, Japan}

\emailAdd{kazumi@azusa.shinshu-u.ac.jp}

\abstract{
We consider matter correlators in the double-scaled SYK (DSSYK)
model.
It turns out that matter correlators have a simple expression 
in terms of the doubled Hilbert space 
$\mathcal{H}\otimes\mathcal{H}$, where 
$\mathcal{H}$ is the Fock space of $q$-deformed oscillator
(also known as the chord Hilbert space).
In this formalism, we find that
the operator which counts the intersection of chords should be
conjugated by certain ``entangler'' and ``disentangler''.  
We explicitly demonstrate this structure for the two- and four-point
functions of matter operators in DSSYK. 
}

\maketitle

\section{Introduction}
To describe a black hole in AdS,
it is useful to consider the doubled (two-sided)
Hilbert space of boundary CFT.
In particular, the eternal black hole in AdS
corresponds to the thermo-field double state \cite{Maldacena:2001kr}
which is closely related to the idea of ER=EPR \cite{Maldacena:2013xja,VanRaamsdonk:2010pw}.
Recently, the doubled Hilbert space in JT gravity
and the double-scaled SYK (DSSYK) model has been extensively studied
in the literature (see e.g. \cite{Kolchmeyer:2023gwa,Penington:2023dql,Berkooz:2022fso,Goel:2018ubv,Lin:2023trc}
and references therein). 

In this paper, we consider matter correlators of DSSYK
in the doubled Hilbert space formalism.
As shown in \cite{Berkooz:2018jqr},
the correlators of DSSYK reduce 
to the counting problem of chord diagrams, which is exactly solved 
in terms of the
$q$-deformed oscillator $A_\pm$.
The Fock space $\cH$ of the $q$-deformed oscillator, also known as the
chord Hilbert space, can be thought of as the Hilbert space of 
bulk gravity theory \cite{Lin:2022rbf}.
It turns out that matter correlators of DSSYK have a simple
expression in the doubled Hilbert space $\cH\otimes\cH$.
We find that the operator which counts the
intersection of chords is conjugated by the ``entangler'' 
$\cE$ and the ``disentangler'' $\cE^{-1}$ 
(see \eqref{eq:bilocal-state2} and \eqref{eq:G4-conj}).
This structure is reminiscent of the tensor network of MERA \cite{Vidal:2007hda,vidal2010entanglement}.

This paper is organized as follows.
In section \ref{sec:review}, we briefly review 
the known result of matter correlators in DSSYK.
In section \ref{sec:double}, we define a mapping of the operator
$X$ on $\cH$ to the state $|X\ket$ in the doubled Hilbert
space $\cH\otimes\cH$
and rewrite the matter correlators as the overlap
$\bra0,0|X\ket$.
In section \ref{sec:corr},
we perform this rewriting explicitly for the two- and four-point 
functions of matter operators.
We find that the intersection-counting operator
is conjugated by the entangler and the disentangler
as in \eqref{eq:bilocal-state2} and \eqref{eq:G4-conj}.
Finally we conclude in section \ref{sec:conclusion} 
with some discussion on the future
problems.
In appendix \ref{app:formula} we summarize some useful formulae
used in the main text.
In appendix \ref{app:diag} we explain the derivation of \eqref{eq:bilocal-diag}.
 In appendix \ref{app:sym} we prove the crossing
symmetry of the $R$-matrix of $U_{q}(\mathfrak{su}(1,1))$.

\section{Review of DSSYK}\label{sec:review}
In this section we briefly review the result of DSSYK in \cite{Berkooz:2018jqr}.
SYK model is defined by the Hamiltonian for 
$N$ Majorana fermions $\psi_i~(i=1,\cdots,N)$
obeying $\{\psi_i,\psi_j\}=2\cob_{i,j}$
with all-to-all $p$-body interaction
\begin{equation}
\begin{aligned}
 H=\ri^{p/2}\sum_{1\leq i_1<\cdots<i_p\leq N}
J_{i_1\cdots i_p}\psi_{i_1}\cdots\psi_{i_p},
\end{aligned} 
\end{equation}
where $J_{i_1\cdots i_p}$ is a random coupling drawn from the Gaussian distribution.
DSSYK is defined by the scaling limit
\begin{equation}
\begin{aligned}
 N,p\to\infty\quad\text{with}\quad \la=\frac{2p^2}{N}:\text{fixed}.
\end{aligned} 
\label{eq:scaling}
\end{equation}
As shown in \cite{Berkooz:2018jqr}, the ensemble average of the moment $\Tr H^k$ 
reduces to a counting problem
of the intersection number of chord diagrams
\begin{equation}
\begin{aligned}
 \bra \Tr H^k\ket_J=\sum_{\text{chord diagrams}}q^{\#(\text{intersections})}
\end{aligned}
\label{eq:moment} 
\end{equation}
with $q=e^{-\la}$.
This counting problem is solved by introducing the transfer matrix $T$
\begin{equation}
\begin{aligned}
 T=\frac{A_{+}+A_{-}}{\rt{1-q}},
\end{aligned} 
\end{equation}
where $A_{\pm}$ denote the 
$q$-deformed oscillator acting on the chord number state $|n\ket$
\begin{equation}
\begin{aligned}
 A_{+}|n\ket=\rt{1-q^{n+1}}|n+1\ket,\qquad
A_{-}|n\ket=\rt{1-q^{n}}|n-1\ket.
\end{aligned} 
\end{equation}
Note that $A_\pm$ satisfy the $q$-deformed
commutation relations
\begin{equation}
\begin{aligned}
 A_-A_+-qA_+A_-&=1-q,\\
A_-A_+-A_+A_-&=(1-q)q^{\h{N}},
\end{aligned} 
\end{equation}
where $\h{N}$ denotes the number operator
\begin{equation}
\begin{aligned}
 \h{N}|n\ket=n|n\ket.
\end{aligned} 
\end{equation}
Then the moment in \eqref{eq:moment} is written as
\begin{equation}
\begin{aligned}
 \bra \Tr H^k\ket_J=\bra 0|T^k|0\ket.
\end{aligned} 
\end{equation} 
The transfer matrix $T$ becomes diagonal in the $\th$-basis
\begin{equation}
\begin{aligned}
 T|\th\ket=E(\th)|\th\ket,\qquad E(\th)=\frac{2\cos\th}{\rt{1-q}},
\end{aligned} 
\end{equation}
and the overlap of $\bra n|$ and $|\th\ket$ is given by the $q$-Hermite polynomial
$H_n(\cos\th|q)$
\begin{equation}
\begin{aligned}
 \bra n|\th\ket=\frac{H_n(\cos\th|q)}{\rt{(q;q)_n}},
\end{aligned} 
\end{equation}
where $(q;q)_n$ denotes the $q$-Pochhammer symbol 
(see appendix \ref{app:formula} for the definition).
$|\th\ket$ and $|n\ket$ are normalized as
\begin{equation}
\begin{aligned}
 \bra \th|\th'\ket&=\frac{2\pi}{\mu(\th)}\cob(\th-\th'),\quad
\bra n|m\ket=\cob_{n,m},\\
\id&=\int_0^\pi\frac{d\th}{2\pi}\mu(\th)|\th\ket\bra\th|=\sum_{n=0}^\infty
 |n\ket\bra n|,
\end{aligned} 
\label{eq:res-uni}
\end{equation}
and the measure factor $\mu(\th)$ is given by
\begin{equation}
\begin{aligned}
 \mu(\th)=(q,e^{\pm2\ri\th};q)_\infty.
\end{aligned} 
\label{eq:mu-th}
\end{equation}

As discussed in \cite{Berkooz:2018jqr}, we can also consider
the matter operator $\cO_\lap$
\begin{equation}
\begin{aligned}
 \cO_\lap=\ri^{s/2}\sum_{1\leq i_1<\cdots< i_s\leq N}K_{i_1\cdots i_s}\psi_{i_1}\cdots\psi_{i_s}
\end{aligned} 
\label{eq:matter-M}
\end{equation}
with a Gaussian random coefficient $K_{i_1\cdots i_s}$
which is drawn independently from the random coupling
$J_{i_1\cdots i_p}$ in the SYK Hamiltonian.
In the double scaling limit \eqref{eq:scaling},
the effect of this operator can be made finite by taking the limit
$s\to\infty$ with $\lap=s/p$ held fixed.
Then the correlator of $\cO_\lap$'s is also written as a  
counting problem of the chord diagrams
\begin{equation}
\begin{aligned}
 \sum_{\text{chord diagrams}}q^{\#(H\text{-}H\,\text{intersections})}
q^{\lap_i\#(H\text{-}\cO_{\lap_i}\,\text{intersections})}
q^{\lap_i\lap_j\#(\cO_{\lap_i}\text{-}\cO_{\lap_j}\,\text{intersections})}.
\end{aligned} 
\label{eq:chord-count}
\end{equation}
Note that there appear two types of chords in this computation: $H$-chords and 
$\cO$-chords coming from the Wick contraction of random couplings 
$J_{i_1\cdots i_p}$ and $K_{i_1\cdots i_s}$, respectively.
The $\cO$-chord is also called matter chord.

Let us consider the bi-local operator
$\wick{\c \cO_\lap e^{-\bt H}\c \cO_\lap}$, where the overline denotes
the Wick contraction of random coupling $K_{i_1\cdots i_s}$.
As shown in \cite{Berkooz:2018jqr}, 
this operator is given by (see also \cite{Okuyama:2022szh})
\begin{equation}
\begin{aligned}
 \wick{\c \cO_\lap e^{-\bt H}\c \cO_\lap}&=
\sum_{n,m,\ell=0}^\infty \frac{(q^{2\lap};q)_\ell}{(q;q)_\ell}
\rt{\frac{(q;q)_{m+\ell}(q;q)_{n+\ell}}{(q;q)_m(q;q)_n}}
|m+\ell\ket\bra m| q^{\lap\h{N}}e^{-\bt T}q^{\lap\h{N}}|n\ket
\bra n+\ell|.
\end{aligned}
\label{eq:bilocal-sum} 
\end{equation}
Using the relation
\begin{equation}
\begin{aligned}
|n\ket=\frac{A_+^n}{\rt{(q;q)_n}}|0\ket,
\qquad A_+^\ell|n\ket=\rt{\frac{(q;q)_{n+\ell}}{(q;q)_n}}|n+\ell\ket,
\end{aligned} 
\label{eq:n-ket}
\end{equation}
\eqref{eq:bilocal-sum} is rewritten as
\begin{equation}
\begin{aligned}
 \wick{\c \cO_\lap e^{-\bt H}\c \cO_\lap}&=
\sum_{\ell=0}^\infty \frac{(q^{2\lap};q)_\ell}{(q;q)_\ell}
A_+^\ell  q^{\lap\h{N}}e^{-\bt T}q^{\lap\h{N}} A_-^\ell.
\end{aligned} 
\label{eq:bilocal}
\end{equation} 
As shown in \cite{Berkooz:2018jqr},
this bi-local operator commutes with $T$ 
\begin{equation}
\begin{aligned}
 \Bigl[T,\wick{\c \cO_\lap e^{-\bt H}\c \cO_\lap}\Bigr]=0.
\end{aligned} 
\label{eq:T-comm}
\end{equation}
The two-point function of matter operator
$\cO_\lap$ is given by
\begin{equation}
\begin{aligned}
 \bra 0|e^{-\bt_2T} \wick{\c \cO_\lap e^{-\bt_1 H}\c \cO_\lap}|0\ket
= \bra 0|e^{-\bt_2T} q^{\lap\h{N}}e^{-\bt_1 T}|0\ket.
\end{aligned} 
\label{eq:2pt}
\end{equation}
Note that only the $\ell=0$ term in \eqref{eq:bilocal}
contributes to the two-point function since $A_-^\ell|0\ket=0$ for $\ell\geq1$.

Similarly, the uncrossed four-point function is given by
\begin{equation}
\begin{aligned}
 &\bra 0|e^{-\bt_4T} \wick{\c \cO_{\lap_2} e^{-\bt_3 H}\c \cO_{\lap_2}}
e^{-\bt_2T}\wick{\c \cO_{\lap_1} e^{-\bt_1 H}\c \cO_{\lap_1}}|0\ket\\
=&\bra 0|\wick{\c \cO_{\lap_2} e^{-\bt_3 H}\c \cO_{\lap_2}}
e^{-(\bt_2+\bt_4)T}\wick{\c \cO_{\lap_1} e^{-\bt_1 H}\c \cO_{\lap_1}}|0\ket\\
=&\bra 0|e^{-\bt_3T}q^{\lap_2\h{N}}
e^{-(\bt_2+\bt_4)T}q^{\lap_1\h{N}}e^{-\bt_1T}|0\ket.
\end{aligned} 
\end{equation}
In the first equality we have used the relation \eqref{eq:T-comm}
and the last equality follows from the fact that 
only the $\ell=0$ term in \eqref{eq:bilocal}
contributes in this computation when sandwiched between $\bra0|$ and $|0\ket$.

The crossed four-point function is given by
\cite{Berkooz:2018jqr}
\begin{equation}
\begin{aligned}
&\bra 0|e^{-\bt_4T} \wick{\c1 \cO_{\lap_2} e^{-\bt_3H} \c2 \cO_{\lap_1} 
e^{-\bt_2H} \c1 \cO_{\lap_2} e^{-\bt_1H} \c2 \cO_{\lap_1}}|0\ket\\
=&\sum_{\ell=0}^\infty 
\frac{(q^{2\lap_2};q)_\ell}{(q;q)_\ell}q^{\lap_1\ell}
\bra 0|e^{-\bt_4T} A_+^\ell
q^{\lap_2\h{N}}e^{-\bt_3T}q^{\lap_1\h{N}}e^{-\bt_2T}q^{\lap_2\h{N}}A_-^\ell
e^{-\bt_1T} |0\ket.
\end{aligned} 
\label{eq:crossed}
\end{equation}
Here we have suppressed the overall 
factor $q^{\lap_1\lap_2}$ coming from the intersection
of the $\cO_{\lap_1}$-chord and the $\cO_{\lap_2}$-chord.

Let us take a closer look at the two-point function \eqref{eq:2pt}.
Inserting the complete set $\{|n\ket\}_{n=0,1,\cdots}$ in  
\eqref{eq:2pt}, the two-point function becomes
\begin{equation}
\begin{aligned}
 \bra 0|e^{-\bt_2T} q^{\lap\h{N}}e^{-\bt_1 T}|0\ket
=\sum_{n=0}^\infty q^{\lap n}
\bra 0|e^{-\bt_2T}|n\ket\bra n|e^{-\bt_1 T}|0\ket.
\end{aligned} 
\end{equation}
As discussed in \cite{Berkooz:2018jqr,Lin:2022rbf}, 
$|n\ket$ represents the state at a constant time-slice
of the bulk geometry with $n$ $H$-chords threading that slice. 
The factor $q^{\lap n}$ comes from the intersection of
matter chord and $n$ $H$-chords.
Thus $q^{\lap\h{N}}$ in \eqref{eq:2pt} can be thought
of as the operator counting the intersection
of $\cO_\lap$-chord and $H$-chords.
This operator $q^{\lap\h{N}}$ plays an important
role in what follows.

\section{Doubled Hilbert space}\label{sec:double}
As we reviewed in the previous section, the matter correlator
of DSSYK takes the form $\bra 0|X|0\ket$,
where $X$ is a linear operator on the chord Hilbert space
$\cH$ spanned by the chord number states $|n\ket~(n=0,1,\cdots)$
\begin{equation}
\begin{aligned}
 \cH=\bigoplus_{n=0}^\infty\mathbb{C}|n\ket.
\end{aligned} 
\end{equation}
In order to study the matter correlators
in DSSYK, it is useful to consider 
the doubled Hilbert space $\cH\otimes \cH$
and regard the operator $X$ as a state
$|X\ket$ in $\cH\otimes \cH$
\begin{equation}
\begin{aligned}
 X\in\text{End}(\cH)~~~\mapsto~~~ |X\ket \in\cH\otimes\cH.
\end{aligned} 
\label{eq:map-X}
\end{equation}
In terms of the basis $\{|n\ket\}_{n=0,1,\cdots}$, this mapping
\eqref{eq:map-X} is given by 
\begin{equation}
\begin{aligned}
 X=\sum_{n,m=0}^\infty|n\ket\bra n|X|m\ket\bra m|
~~~\mapsto~~~ |X\ket=\sum_{n,m=0}^\infty|n,m\ket \bra n|X|m\ket,
\end{aligned} 
\label{eq:map-X-comp}
\end{equation}
where $|n,m\ket$ is the natural basis of $\cH\otimes \cH$
\begin{equation}
\begin{aligned}
 |n,m\ket:=|n\ket\otimes |m\ket.
\end{aligned} 
\label{eq:nm-def}
\end{equation}
In particular, the identity operator $\id$ corresponds to the state
\begin{equation}
\begin{aligned}
 |\id\ket=\sum_{n=0}^\infty |n,n\ket=\cE|0,0\ket,
\end{aligned} 
\label{eq:id-state}
\end{equation}
where $\cE$ is given by (see \eqref{eq:n-ket} and \eqref{eq:sum1})
\begin{equation}
\begin{aligned}
 \cE=\sum_{n=0}^\infty \frac{A_+^n\otimes A_+^n}{(q;q)_n}
=\frac{1}{(A_+\otimes A_+;q)_\infty}.
\end{aligned} 
\label{eq:cE}
\end{equation}
Note that the state $|\id\ket$ is the maximally entangled state 
and the operator $\cE$ generates the entanglement when acting on the pure state
$|0,0\ket$.
Similarly, the operator $q^{\lap\h{N}}$ corresponds to the state
$|q^{\lap\h{N}}\ket$
\begin{equation}
\begin{aligned}
 \bigl|q^{\lap\h{N}}\bigr\ket=\sum_{n=0}^\infty q^{\lap n}|n,n\ket=\cE_\lap|0,0\ket
\end{aligned} 
\label{eq:state-lap}
\end{equation}
with
\begin{equation}
\begin{aligned}
 \cE_\lap=\frac{1}{(q^\lap A_+\otimes A_+;q)_\infty}.
\end{aligned} 
\label{eq:cE-lap}
\end{equation}

Note that we can append and/or prepend strings of operators
as \footnote{A similar construction is discussed in \cite{Stanford-talk}.}
\begin{equation}
\begin{aligned}
 |XYZ\ket=(X\otimes {}^tZ)|Y\ket
\end{aligned} 
\label{eq:append}
\end{equation}
where $X,Y,Z\in\text{End}(\cH)$ and ${}^tZ$ denotes the transpose of $Z$
\begin{equation}
\begin{aligned}
 \bra n|{}^tZ|m\ket=\bra m|Z|n\ket.
\end{aligned} 
\end{equation}
We should stress that we do not take the complex conjugation of $Z$ 
on the right hand side of \eqref{eq:append};
we simply reverse the order of multiplication and take the transpose of $Z$
in \eqref{eq:append}.

As an example of \eqref{eq:append}, let us consider the relation
\begin{equation}
\begin{aligned}
A_- q^{\lap \h{N}}=q^{\lap\h{N}}q^{\lap}A_-.
\end{aligned} 
\end{equation}
Using ${}^t A_\pm=A_\mp$ and  \eqref{eq:append}, we find
\footnote{The state $\bigl|q^{\lap\h{N}}\bigr\ket$ in \eqref{eq:state-lap}
is reminiscent
of the boundary state $|B_a\ket$ of the end of the world brane
\cite{Okuyama:2023byh}
\begin{equation}
\begin{aligned}
 |B_a\ket=\frac{1}{(aA_+;q)_\infty}|0\ket.
\end{aligned} 
\end{equation}
As shown in \cite{Okuyama:2023byh}, the boundary state $|B_a\ket$ is a coherent state
of the $q$-deformed oscillator
\begin{equation}
\begin{aligned}
 A_-|B_a\ket=a|B_a\ket,
\end{aligned} 
\end{equation}
where the parameter $a$ is related to the tension of the brane.
}
\begin{equation}
\begin{aligned}
 (A_-\otimes \id)\bigl|q^{\lap\h{N}}\bigr\ket=(\id\otimes q^{\lap}A_+)\bigl|q^{\lap\h{N}}\bigr\ket.
\end{aligned} 
\label{eq:coh1}
\end{equation}
We can also show that
\begin{equation}
\begin{aligned}
 (\id\otimes A_-)\bigl|q^{\lap\h{N}}\bigr\ket=(q^{\lap}A_+\otimes\id)\bigl|q^{\lap\h{N}}\bigr\ket.
\end{aligned} 
\label{eq:coh2}
\end{equation}

From \eqref{eq:res-uni}, the state $|\id\ket$ is written 
in terms of the $|\th\ket$-basis as
\begin{equation}
\begin{aligned}
 |\id\ket=\int_0^\pi\frac{d\th}{2\pi}\mu(\th)|\th,\th\ket,
\end{aligned} 
\end{equation}
and the state
corresponding to 
the operator $e^{-\bt T}$ is given by
\begin{equation}
\begin{aligned}
 |e^{-\bt T}\ket=(e^{-\hf\bt T}\otimes e^{-\hf\bt T})|\id\ket=
\int_0^\pi\frac{d\th}{2\pi}\mu(\th)e^{-\bt E(\th)}|\th,\th\ket.
\end{aligned} 
\label{eq:TFD}
\end{equation}
This state $|e^{-\bt T}\ket$ is known as the thermo-field double state.

\section{Matter correlators in the doubled Hilbert space formalism}\label{sec:corr}
In this section, we consider matter correlators of DSSYK
in the doubled Hilbert space formalism.
In general, the matter correlator of DSSYK takes 
the form $\bra 0|X|0\ket$
with some operator $X\in\text{End}(\cH)$.
In the doubled Hilbert space formalism, $\bra 0|X|0\ket$ is expressed as
\begin{equation}
\begin{aligned}
 \bra 0|X|0\ket=\bra 0,0|X\ket.
\end{aligned} 
\label{eq:corr-00X}
\end{equation} 

\subsection{Two-point function}\label{sec:2pt}
Let us first consider the bi-local operator in \eqref{eq:bilocal},
which is the basic building block of the two-point function and
the uncrossed four-point function.
The state $|\wick{\c \cO_\lap e^{-\bt H}\c \cO_\lap}\ket$
corresponding to the operator in \eqref{eq:bilocal}
is given by
\begin{equation}
\begin{aligned}
 |\wick{\c \cO_\lap e^{-\bt H}\c \cO_\lap}\ket&=\sum_{\ell=0}^\infty
\frac{(q^{2\lap};q)_\ell}{(q;q)_\ell}(A_+^\ell\otimes
A_+^\ell) (q^{\lap\h{N}}\otimes q^{\lap\h{N}})|e^{-\bt T}\ket\\
&=\frac{(q^{2\lap}A_+\otimes A_+;q)_\infty}{(A_+\otimes A_+;q)_\infty} 
(q^{\lap\h{N}}\otimes q^{\lap\h{N}})|e^{-\bt T}\ket,
\end{aligned} 
\label{eq:bilocal-state}
\end{equation}
where we used the summation formula in \eqref{eq:sum2}.
Using the relation 
\begin{equation}
\begin{aligned}
 q^{\lap\h{N}}A_+=q^{\lap}A_+ q^{\lap\h{N}},
\end{aligned}
\label{eq:A-shift} 
\end{equation}
\eqref{eq:bilocal-state} is rewritten as
\begin{equation}
\begin{aligned}
 |\wick{\c \cO_\lap e^{-\bt H}\c \cO_\lap}\ket&=
\cE (q^{\lap\h{N}}\otimes q^{\lap\h{N}})\cE^{-1}|e^{-\bt T}\ket,
\end{aligned} 
\label{eq:bilocal-state2}
\end{equation}
where $\cE$ is defined in \eqref{eq:cE}.
The appearance of the operator $q^{\lap\h{N}}\otimes q^{\lap\h{N}}$ 
in \eqref{eq:bilocal-state2} is natural since it counts the number of
intersections between the $H$-chord and the matter chord.
The important point is that this operator $q^{\lap\h{N}}\otimes q^{\lap\h{N}}$ 
should be conjugated by $\cE$
\begin{equation}
\begin{aligned}
 q^{\lap\h{N}}\otimes q^{\lap\h{N}}\to 
\cE(q^{\lap\h{N}}\otimes q^{\lap\h{N}})\cE^{-1}.
\end{aligned} 
\label{eq:conj-cE}
\end{equation}
This conjugation guarantees that the $\bt\to0$ limit of the state
\eqref{eq:bilocal-state2} reduces to $|\id\ket$
in \eqref{eq:id-state}
\begin{equation}
\begin{aligned}
 \lim_{\bt\to0}|\wick{\c \cO_\lap e^{-\bt H}\c \cO_\lap}\ket&=
\cE (q^{\lap\h{N}}\otimes q^{\lap\h{N}})\cE^{-1}|\id\ket\\
&=\cE (q^{\lap\h{N}}\otimes q^{\lap\h{N}})|0,0\ket\\
&=\cE |0,0\ket\\
&=|\id\ket.
\end{aligned} 
\end{equation}
In other words, the conjugation \eqref{eq:conj-cE} is necessary for
the following operator identity to hold 
\footnote{See also footnote 1 in \cite{Okuyama:2022szh}.}
\begin{equation}
\begin{aligned}
 \wick{\c \cO_\lap \c \cO_\lap}=\id.
\end{aligned} 
\end{equation} 
Following the language of tensor networks, we call $\cE$ and
$\cE^{-1}$ as ``entangler'' and ``disentangler'', respectively.
Our result \eqref{eq:bilocal-state2} shows that we have to insert the disentangler
$\cE^{-1}$ before acting the intersection-counting
operator
$q^{\lap\h{N}}\otimes q^{\lap\h{N}}$.
In the context of MERA \cite{Vidal:2007hda,vidal2010entanglement}, 
disentanglers are usually assumed to be
unitary operators. However, 
our $\cE$ and $\cE^{-1}$ are not unitary.
Thus, \eqref{eq:conj-cE} is a similarity transformation, 
not a unitary transformation.

From the time-translation invariance 
\eqref{eq:T-comm} of the bi-local operator 
$\wick{\c \cO_\lap e^{-\bt H}\c \cO_\lap}$, it follows that 
the state $|\wick{\c \cO_\lap e^{-\bt H}\c \cO_\lap}\ket$ in 
\eqref{eq:bilocal-state2}  is diagonal in the $|\th\ket$-basis
\begin{equation}
\begin{aligned}
 |\wick{\c \cO_\lap e^{-\bt H}\c \cO_\lap}\ket=\int_0^\pi\frac{d\th}{2\pi}\mu(\th)
|\th,\th\ket\bra\th|q^{\lap\h{N}}e^{-\bt T}|0\ket.
\end{aligned} 
\label{eq:bilocal-diag}
\end{equation}
See appendix \ref{app:diag} for the derivation of this expression.

\subsection{Crossed four-point function}\label{sec:4pt}
Next, let us consider the crossed four-point function \eqref{eq:crossed}
\begin{equation}
\begin{aligned}
 G_4:=\sum_{\ell=0}^\infty 
\frac{(q^{2\lap_2};q)_\ell}{(q;q)_\ell}q^{\lap_1\ell}
\bra 0|e^{-\bt_4T} A_+^\ell
q^{\lap_2\h{N}}e^{-\bt_3T}q^{\lap_1\h{N}}e^{-\bt_2T}q^{\lap_2\h{N}}A_-^\ell
e^{-\bt_1T} |0\ket.
\end{aligned} 
\end{equation}
In the doubled Hilbert space formalism, this is written as
\begin{equation}
\begin{aligned}
 G_4=
\sum_{\ell=0}^\infty 
\frac{(q^{2\lap_2};q)_\ell}{(q;q)_\ell}q^{\lap_1\ell}
\bra 0,0|(e^{-\bt_4T}\otimes e^{-\bt_1T})(A_+^\ell\otimes A_+^\ell)
(q^{\lap_2\h{N}}\otimes q^{\lap_2\h{N}})
(e^{-\bt_3T}\otimes e^{-\bt_2T})|q^{\lap_1\h{N}}\ket.
\end{aligned} 
\label{eq:G4-sum}
\end{equation}
Using \eqref{eq:sum2} and \eqref{eq:A-shift}, one can show that
\eqref{eq:G4-sum} is written as
\begin{equation}
\begin{aligned}
 G_4=\bra 0,0|(e^{\bt_4T}\otimes e^{\bt_1T})\cE_{\lap_1}
(q^{\lap_2\h{N}}\otimes q^{\lap_2\h{N}})
(\cE_{\lap_1})^{-1}(e^{\bt_3T}\otimes e^{\bt_2T})|q^{\lap_1\h{N}}\ket,
\end{aligned} 
\label{eq:G4-conj}
\end{equation}
where $\cE_{\lap_1}$ is defined in \eqref{eq:cE-lap}.
Again, the operator $q^{\lap_2\h{N}}\otimes q^{\lap_2\h{N}}$
is conjugated by $\cE_{\lap_1}$ in \eqref{eq:G4-conj}; 
$\cE_{\lap_1}$ and $(\cE_{\lap_1})^{-1}$ can be thought of as the entangler 
and the disentangler associated with the state
$|q^{\lap_1\h{N}}\ket=\cE_{\lap_1}|0,0\ket$.
$G_4$ in \eqref{eq:G4-conj} is schematically depicted as
\begin{equation}
\begin{aligned}
 \begin{tikzpicture}[scale=1]
\draw (0,0) circle [radius=2];
\draw[blue,thick] (0,-2)--(0,2); 
\draw[red,thick] (-2,0)--(2,0); 
\draw (2.1,0) node [right]{$|q^{\lap_1\h{N}}\ket$};
\draw (-2.1,0) node [left]{$\bra0,0|$};
\draw (1.45,-1.45) node [right]{$\bt_2$};
\draw (1.45,1.45) node [right]{$\bt_3$};
\draw (-1.45,1.45) node [left]{$\bt_4$};
\draw (-1.45,-1.45) node [left]{$\bt_1$};
\end{tikzpicture}
\end{aligned}~,
\label{eq:G4-pic}
\end{equation}
where the red line and the blue line correspond to the $\cO_{\lap_1}$-chord
and the $\cO_{\lap_2}$-chord, respectively.
In this picture, the bra and the ket are treated asymmetrically and
some of the symmetries of $G_4$ are not
manifest in our representation \eqref{eq:G4-conj}.
In particular,
the crossing symmetry $(12)\leftrightarrow(34)$ of $G_4$ is not
manifest in \eqref{eq:G4-conj}.

The crossing symmetry (or the exchange of bra and ket)
of $G_4$ can be  seen as follows
(see appendix \ref{app:sym} for the details).
Inserting the resolution of identity $\id$ in \eqref{eq:res-uni},
$G_4$ in \eqref{eq:G4-conj}
is written as
\begin{equation}
\begin{aligned}
 G_4=\int\prod_{k=1}^4\frac{d\th_k}{2\pi}\mu(\th_k)e^{-\bt_k E(\th_k)}R(\th_1,\th_2,\th_3,\th_4)
\end{aligned} 
\end{equation}
with
\begin{equation}
\begin{aligned}
 R(\th_1,\th_2,\th_3,\th_4)=\bra \th_4,\th_1|
\cE_{\lap_1}(q^{\lap_2\h{N}}\otimes q^{\lap_2\h{N}})(\cE_{\lap_1})^{-1}
|\th_3,\th_2\ket\bra\th_3|q^{\lap_1\h{N}}|\th_2\ket.
\end{aligned} 
\label{eq:R1234}
\end{equation}
This $R(\th_1,\th_2,\th_3,\th_4)$ is proportional to the $R$-matrix of the quantum
group $U_q(\mathfrak{su}(1,1))$, which is
written in terms of the basic hypergeometric series
${}_8W_7$.
Using the Bailey transformation \eqref{eq:bailey}
of ${}_8W_7$, one can show that
$R(\th_1,\th_2,\th_3,\th_4)$ in \eqref{eq:R1234} is invariant under the crossing symmetry
$(12)\leftrightarrow(34)$
\begin{equation}
\begin{aligned}
 R(\th_1,\th_2,\th_3,\th_4)=R(\th_3,\th_4,\th_1,\th_2),
\end{aligned}
\label{eq:cross-R} 
\end{equation}
which implies that $G_4$ is invariant under $(\bt_1\bt_2)\leftrightarrow
(\bt_3\bt_4)$.
This symmetry of $G_4$ is schematically depicted as
\begin{equation}
\begin{aligned}
 \begin{tikzpicture}[scale=1]
\draw (0,0) circle [radius=2];
\draw[blue,thick] (0,-2)--(0,2); 
\draw[red,thick] (-2,0)--(2,0); 
\draw (2.1,0) node [right]{$|q^{\lap_1\h{N}}\ket$};
\draw (-2.1,0) node [left]{$\bra0,0|$};
\draw (1.45,-1.45) node [right]{$\bt_2$};
\draw (1.45,1.45) node [right]{$\bt_3$};
\draw (-1.45,1.45) node [left]{$\bt_4$};
\draw (-1.45,-1.45) node [left]{$\bt_1$};
\draw (3.4,0) node [right]{$=$};
\draw (7,0) circle [radius=2];
\draw[blue,thick] (7,-2)--(7,2); 
\draw[red,thick] (9,0)--(5,0); 
\draw (9.1,0) node [right]{$|q^{\lap_1\h{N}}\ket$};
\draw (4.9,0) node [left]{$\bra0,0|$};
\draw (8.45,-1.45) node [right]{$\bt_4$};
\draw (8.45,1.45) node [right]{$\bt_1$};
\draw (5.55,1.45) node [left]{$\bt_2$};
\draw (5.55,-1.45) node [left]{$\bt_3$};
\end{tikzpicture}
\end{aligned}~.
\label{eq:G4-cross}
\end{equation}

\section{Conclusion and outlook}\label{sec:conclusion}
In this paper we have studied the matter correlators of DSSYK in the doubled Hilbert
space formalism.
In our formalism, a matter correlator of the form $\bra 0|X|0\ket$
is expressed as the overlap between $\bra 0,0|$ and the state
$|X\ket\in\cH\otimes\cH$ corresponding to the operator $X$, where
the relation between $X$ and $|X\ket$ is given by \eqref{eq:map-X-comp}.
We find that the intersection-counting operator $q^{\lap\h{N}}\otimes q^{\lap\h{N}}$
should be conjugated by the entangler $\cE$ and the disentangler $\cE^{-1}$
as in \eqref{eq:bilocal-state2} (or the entangler $\cE_{\lap_1}$ and 
disentangler $(\cE_{\lap_1})^{-1}$ in the case of 
crossed four-point function \eqref{eq:G4-conj}).
In our representation of a matter correlator $\bra0,0|X\ket$ \eqref{eq:corr-00X}, 
the bra and the ket are treated 
asymmetrically and hence some of the symmetries of the correlators
are not manifest.
Nevertheless, the bra-ket exchange symmetry (or crossing symmetry)
of the four-point 
function \eqref{eq:G4-cross}
can be shown rather non-trivially by using the Bailey transformation
\eqref{eq:bailey} of ${}_8W_7$.

We should stress that our formalism is different from that in \cite{Lin:2023trc}.
The authors of \cite{Lin:2023trc} introduced 
the two-sided chord Hilbert space in the presence of the matter operator, 
spanned by the states $\{|n_L,n_R\ket\}$ where $n_L$ and $n_R$
denote the number of $H$-chords to the left and right of the matter chord.
Our $|n,m\ket$ in \eqref{eq:nm-def} is not equal to 
$|n_L,n_R\ket$ in \cite{Lin:2023trc}.
According to the discussion in \cite{Stanford-talk}, 
our $|n,m\ket$ can be expanded as a linear combination 
of $|n_L,n_R\ket$ in \cite{Lin:2023trc}.
It would be interesting to find a precise relation between our
 $|n,m\ket$ and $|n_L,n_R\ket$ in \cite{Lin:2023trc}.

The construction of the two-sided chord Hilbert space in \cite{Lin:2023trc}
is based on a picture of cutting open the ``bulk path integral''.
On the other hand,
our formalism is based on a honest, direct rewriting of the known result
of matter correlators in \cite{Berkooz:2018jqr}.
At present we do not understand clearly how these two approaches are related.
In particular, in our formalism we do not need to introduce the co-product
of $q$-deformed oscillator $A_\pm$, which played an important role in
the discussion of symmetry algebra in \cite{Lin:2023trc}.
Perhaps, \eqref{eq:coh1} and \eqref{eq:coh2}
might be a good starting point to consider the relationship 
between the two approaches. 
We leave this as an interesting future problem.

\acknowledgments
This work was supported
in part by JSPS Grant-in-Aid for Transformative Research Areas (A) 
``Extreme Universe'' 21H05187 and JSPS KAKENHI Grant 22K03594.

\appendix
\section{Useful formulae}\label{app:formula}
In this appendix, we summarize some useful formulae used in the main text.
The $q$-Pochhammer symbol is defined by
\begin{equation}
\begin{aligned}
 (a;q)_n=\prod_{k=0}^{n-1}(1-aq^k),\qquad
(a_1,\cdots,a_s;q)_n=\prod_{i=1}^s(a_i;q)_n.
\end{aligned} 
\end{equation}
The following summation formulae play an important role in this paper:
\begin{align}
\label{eq:sum1}
 \sum_{n=0}^\infty\frac{t^n}{(q;q)_n}&=\frac{1}{(t;q)_\infty},\\
\label{eq:sum2}
 \sum_{n=0}^\infty\frac{(a;q)_n}{(q;q)_n}t^n&=\frac{(ta;q)_\infty}{(t;q)_\infty}.
\end{align}

The $q$-Hermite polynomial $H_n(x|q)$ is defined by
the recursion relation
\begin{equation}
\begin{aligned}
 2xH_n(x|q)=H_{n+1}(x|q)+(1-q^n)H_{n-1}(x|q),
\end{aligned} 
\end{equation}
with the initial condition $H_{-1}=0,H_0=1$.
In the computation of matter correlators, we need the following formula for the Poisson kernel of the $q$-Hermite polynomials
\begin{equation}
\begin{aligned}
 \bra\th_1|t^{\h{N}}|\th_2\ket&=\sum_{n=0}^\infty 
t^n\bra\th_1|n\ket\bra n|\th_2\ket\\
&=\sum_{n=0}^\infty\frac{t^n}{(q;q)_n}H_n(\cos\th_1|q)
H_n(\cos\th_2|q)\\
&=\frac{(t^2;q)_\infty}{(te^{\ri(\pm\th_1\pm\th_2)};q)_\infty}.
\end{aligned} 
\label{eq:poisson-H}
\end{equation}

Al-Salam-Chihara polynomial $Q_n(x|a,b,q)$ is defined by
the recursion relation
\begin{equation}
\begin{aligned}
 2xQ_n=Q_{n+1}+(a+b)q^nQ_n+(1-q^n)(1-abq^{n-1})Q_{n-1}
\end{aligned} 
\end{equation}
with the initial condition $Q_{-1}=0,Q_0=1$.
Using the summation formula in \cite{szablowski2013q}, we find
that the matrix element of $A_+^\ell q^{\lap\h{N}}$ is given
by the Al-Salam-Chihara polynomial
\begin{equation}
\begin{aligned}
 \bra \th_1|A_+^\ell t^{\h{N}}|\th_2\ket&=\sum_{n=0}^\infty 
t^n\bra\th_1|A_+^\ell|n\ket\bra n|\th_2\ket\\
&=\sum_{n=0}^\infty\frac{t^n}{(q;q)_n}
H_{n+\ell}(\cos\th_1|q)H_{n}(\cos\th_2|q)\\
&=\bra \th_1|t^{\h{N}}|\th_2\ket
\frac{Q_\ell(\cos\th_1|te^{\pm\ri\th_2},q)}{(t^2;q)_\ell}.
\end{aligned} 
\label{eq:A+Q}
\end{equation}

\section{Derivation of \eqref{eq:bilocal-diag}}\label{app:diag}
In this appendix, we derive the relation \eqref{eq:bilocal-diag}.
To this end, let us consider the overlap of $\bra\th_1,\th_2|$ and
the state $|\wick{\c\cO_{\lap} e^{-\bt H}\c \cO_{\lap}}\ket$ in 
\eqref{eq:bilocal-state}
\begin{equation}
\begin{aligned}
 \bra\th_1,\th_2|\wick{\c\cO_{\lap} e^{-\bt H}\c \cO_{\lap}}\ket&=
\sum_{\ell=0}^\infty \frac{(q^{2\lap};q)_\ell}{(q;q)_\ell}
\int_0^\pi\frac{d\th}{2\pi}\mu(\th)e^{-\bt E(\th)}
\bra\th_1|A_+^\ell q^{\lap\h{N}}|\th\ket
\bra\th_2|A_+^\ell q^{\lap\h{N}}|\th\ket,
\end{aligned} 
\label{eq:sumQ12}
\end{equation}
where we used the integral form of the thermo-field
double state $|e^{-\bt T}\ket$ in \eqref{eq:TFD}.
Plugging \eqref{eq:A+Q} into \eqref{eq:sumQ12}, we find
\begin{equation}
\begin{aligned}
 \bra\th_1,\th_2|\wick{\c\cO_{\lap} e^{-\bt H}\c \cO_{\lap}}\ket&=
\int_0^\pi\frac{d\th}{2\pi}\mu(\th)e^{-\bt E(\th)}\bra\th_1|q^{\lap\h{N}}|\th\ket
\bra\th_2|q^{\lap\h{N}}|\th\ket\\
&\times \sum_{\ell=0}^\infty \frac{Q_\ell(\cos\th_1|q^{\lap}e^{\pm\ri\th},q)
Q_\ell(\cos\th_2|q^{\lap}e^{\pm\ri\th},q)}{(q,q^{2\lap};q)_\ell}.
\end{aligned} 
\label{eq:sumQ12-2}
\end{equation}
Using the summation formula \cite{askey1996general}
\begin{equation}
\begin{aligned}
 \sum_{\ell=0}^\infty \frac{Q_\ell(\cos\th_1|q^{\lap}e^{\pm\ri\th},q)
Q_\ell(\cos\th_2|q^{\lap}e^{\pm\ri\th},q)}{(q,q^{2\lap};q)_\ell}
=\frac{\bra\th_1|\th_2\ket}{\bra\th_1|q^{\lap\h{N}}|\th\ket},
\end{aligned} 
\end{equation}
\eqref{eq:sumQ12-2} becomes
\begin{equation}
\begin{aligned}
 \bra\th_1,\th_2|\wick{\c\cO_{\lap} e^{-\bt H}\c \cO_{\lap}}\ket&=
\bra\th_1|\th_2\ket\int_0^\pi\frac{d\th}{2\pi}\mu(\th)e^{-\bt E(\th)}
\bra\th_1|q^{\lap\h{N}}|\th\ket\\
&=\bra\th_1|\th_2\ket\int_0^\pi\frac{d\th}{2\pi}\mu(\th)
\bra\th_1|q^{\lap\h{N}}e^{-\bt T}|\th\ket\\
&=\bra\th_1|\th_2\ket\bra\th_1|q^{\lap\h{N}}e^{-\bt T}|0\ket.
\end{aligned} 
\label{eq:overlap}
\end{equation}
In the last equality we used the relation
\begin{equation}
\begin{aligned}
 \int_0^\pi\frac{d\th}{2\pi}\mu(\th)|\th\ket=|0\ket,
\end{aligned} 
\end{equation}
where $|0\ket$ on the right hand side stands for $|n=0\ket$.
One can easily see that \eqref{eq:overlap} is equivalent to
our desired relation \eqref{eq:bilocal-diag}. This completes the proof of 
\eqref{eq:bilocal-diag}. 

\section{Crossing symmetry of $R(\th_1,\th_2,\th_3,\th_4)$}\label{app:sym}
In this appendix, we prove the crossing symmetry of $R(\th_1,\th_2,\th_3,\th_4)$
in \eqref{eq:cross-R}.
Using \eqref{eq:A+Q}, $R(\th_1,\th_2,\th_3,\th_4)$ 
in \eqref{eq:R1234} is written as
\begin{equation}
\begin{aligned}
 R(\th_1,\th_2,\th_3,\th_4)&=
\bra \th_4,\th_1|\left(\sum_{\ell=0}^\infty\frac{(q^{2\lap_2};q)_\ell}{(q;q)_\ell}
q^{\lap_1\ell}A_+^\ell\otimes A_+^\ell\right)(q^{\lap_2\h{N}}\otimes q^{\lap_2\h{N}})
|\th_3,\th_2\ket\bra\th_3|q^{\lap_1\h{N}}|\th_2\ket\\
&=\bra\th_3|q^{\lap_1\h{N}}|\th_2\ket
\bra\th_4|q^{\lap_2\h{N}}|\th_3\ket
\bra\th_1|q^{\lap_2\h{N}}|\th_2\ket\\&\quad \times
\sum_{\ell=0}^\infty\frac{q^{\lap_1\ell}}{(q^{2\lap_2},q;q)_\ell}
Q_\ell(\cos\th_4|q^{\lap_2}e^{\pm\ri\th_3};q)
Q_\ell(\cos\th_1|q^{\lap_2}e^{\pm\ri\th_2};q)\\
&=
\bra\th_4|q^{\lap_2\h{N}}|\th_3\ket\bra\th_3|q^{\lap_1\h{N}}|\th_2\ket
\bra\th_2|q^{\lap_2\h{N}}|\th_1\ket
\bra\th_1|q^{\lap_1\h{N}}|\th_4\ket\\&\times
\frac{(q^{\lap_1}e^{-\ri(\th_2+\th_3)},q^{\lap_1+\lap_2}e^{\ri(\th_2\pm\th_4)},
q^{\lap_1+\lap_2}e^{\ri(\th_3\pm\th_1)};q)_\infty}{(q^{2\lap_1},
q^{\lap_1+2\lap_2}e^{\ri(\th_2+\th_3)};q)_\infty}\\
&\times{}_8W_7(q^{-1+\lap_1+2\lap_2}e^{\ri(\th_2+\th_3)};
q^{\lap_1}e^{\ri(\th_2+\th_3)},q^{\lap_2}e^{\ri(\th_3\pm\th_4)},
q^{\lap_2}e^{\ri(\th_2\pm\th_1)};q,
q^{\lap_1}e^{-\ri(\th_2+\th_3)}).
\end{aligned} 
\end{equation}
In the last step, we have used the Poisson kernel of the Al-Salam-Chihara
polynomials \cite{askey1996general}
\begin{equation}
\begin{aligned}
&\frac{1}{\bra\th_1|t^{\h{N}}|\th_4\ket} \sum_{\ell=0}^\infty
\frac{t^\ell}{(ab,q;q)_\ell}Q_\ell(\cos\th_4|a,b;q)
Q_\ell(\cos\th_1|\al,\bt;q)\\
=&
\frac{(\bt a^{-1}t,\al te^{\pm\ri\th_4},
ate^{\pm\ri\th_1};q)_\infty}
{(t^2,a\al t;q)_\infty}
{}_8W_7(q^{-1}a\al t;\al b^{-1}t,ae^{\pm\ri\th_4},\al
e^{\pm\ri\th_1}
;q,\bt a^{-1}t)
\end{aligned} 
\label{eq:R-8W7}
\end{equation}
where $ab=\al\bt$ and
the well-poised basic hypergeometric series
${}_8W_7$
is defined by
\footnote{The basic hypergeometric series
\begin{equation}
\begin{aligned}
 \sum_{n=0}^\infty \frac{(a,\al_1,\cdots,\al_s;q)_n}{(q,\bt_1,\cdots,\bt_s;q)_n}z^n
\end{aligned} 
\end{equation}
is called well-poised when $aq=\al_1\bt_1=\cdots=\al_s\bt_s$.
${}_8W_7(a;b,c,d,e,f;q,z)$ is called very well-poised when $z=\la q/ef$.
}
\begin{equation}
\begin{aligned}
  {}_8W_7(a;b,c,d,e,f;q,z)=\sum_{n=0}^\infty \frac{(a,\pm qa^{\hf},b,c,d,e,f;q)_n}
{(q,\pm a^{\hf}, qa/b,qa/c,qa/d,qa/e,qa/f;q)_n}z^n.
\end{aligned} 
\end{equation}

The crossing symmetry of $R(\th_1,\th_2,\th_3,\th_4)$ 
in \eqref{eq:R-8W7} can be shown by using 
the Bailey transform of ${}_8W_7$ (see e.g. \cite{gasper1995lecture})
\begin{equation}
\begin{aligned}
 {}_8W_7\Bigl(a;b,c,d,e,f;q,\frac{\la q}{ef}\Bigr)
=\frac{\Bigl(aq,\frac{aq}{ef},\frac{\la q}{e},\frac{\la q}{f};q\Bigr)_\infty}
{\Bigl(\la q,\frac{\la q}{ef},\frac{a q}{e},\frac{a q}{f};q\Bigr)_\infty}
{}_8W_7\Bigl(\la;\tilde{b},\tilde{c},\tilde{d},e,f;q,\frac{aq}{ef}\Bigr),
\end{aligned} 
\label{eq:bailey}
\end{equation}
where
\begin{equation}
\begin{aligned}
 \la=\frac{qa^2}{bcd},\quad \tilde{b}=\frac{\la b}{a},\quad
\tilde{c}=\frac{\la c}{a},\quad
\tilde{d}=\frac{\la d}{a}.
\end{aligned} 
\label{eq:def-la}
\end{equation}
Note that the dual version of the first relation of \eqref{eq:def-la}
is given by
\begin{equation}
\begin{aligned}
 a=\frac{q\la^2}{\tilde{b}\tilde{c}\tilde{d}}.
\end{aligned} 
\end{equation}
Thus the transformation of the parameters $(a,b,c,d)\to(\la,\tilde{b},\tilde{c},\tilde{d})$ in \eqref{eq:bailey} is 
a $\mathbb{Z}_2$ involution.\footnote{More generally, 
${}_8W_7$ has a symmetry of 
$W(D_5)$, the Weyl group of the root system $D_5$ \cite{van1999invariance}.}
We can apply the Bailey transformation \eqref{eq:bailey} to our case
\eqref{eq:R-8W7} by setting
\begin{equation}
\begin{aligned}
 a&=q^{-1+\lap_1+2\lap_2}e^{\ri(\th_2+\th_3)},\\
b&=q^{\lap_1}e^{\ri(\th_2+\th_3)},\\
c&=q^{\lap_2}e^{\ri(\th_2-\th_1)},\\
d&=q^{\lap_2}e^{\ri(\th_3-\th_4)},\\
e&=q^{\lap_2}e^{\ri(\th_1+\th_2)},\\
f&=q^{\lap_2}e^{\ri(\th_3+\th_4)}.
\end{aligned}
\label{eq:orig-param} 
\end{equation}
Then the dual parameters are given by
\begin{equation}
\begin{aligned}
\la&=q^{-1+\lap_1+2\lap_2}e^{\ri(\th_1+\th_4)},\\
 \tilde{b}&=q^{\lap_1}e^{\ri(\th_1+\th_4)},\\
\tilde{c}&=q^{\lap_2}e^{\ri(\th_4-\th_3)},\\
\tilde{d}&=q^{\lap_2}e^{\ri(\th_1-\th_2)}.
\end{aligned} 
\label{eq:dual-param}
\end{equation}
We can see that the mapping from $(a,b,c,d)$ 
in \eqref{eq:orig-param} to $(\la,\tilde{b},\tilde{c},\tilde{d})$
in \eqref{eq:dual-param} 
corresponds to the crossing symmetry $(12)\leftrightarrow(34)$.
We can also check that the prefactor of ${}_8W_7$ in \eqref{eq:R-8W7}
is correctly transformed
under the Bailey transformation \eqref{eq:bailey}.
Finally we find that $R(\th_1,\th_2,\th_3,\th_4)$
in \eqref{eq:R-8W7} is invariant under the crossing symmetry \eqref{eq:cross-R}.

\bibliography{paper}
\bibliographystyle{utphys}

\end{document}